\documentclass[]{article}

\usepackage[utf8]{inputenc} 

\usepackage{geometry} 
\usepackage{graphicx}

\usepackage{booktabs} 
\usepackage{array} 
\usepackage{paralist} 
\usepackage{verbatim} 
\usepackage{subfig} 

\usepackage{fancyhdr} 
\usepackage{sectsty}
\allsectionsfont{\sffamily\mdseries\upshape} 

\usepackage[nottoc,notlof,notlot]{tocbibind} 
\usepackage[titles,subfigure]{tocloft} 

\title{A 1D kinetic model for cosmic microwave background comptonization}
\author{A. Sandoval-Villalbazo \\
Department of Physics and Mathematics \\  U.\ Iberoamericana \\
	\\Prolongaci\'on Paseo de la Reforma 880   Lomas de Santa Fe 01219 \\ M\'exico City, M\'exico}
\begin{document}

\maketitle

\begin{abstract}
This work presents a novel derivation of the expressions that  describe the distortions of the cosmic microwave background curve due to the interactions between photons and the electrons present in dilute ionized systems. In this approach, a simplified   one-dimensional evolution  equation for the photon number occupation is applied   to describe the mentioned  interaction. This methodology emphasizes the physical features of the Sunyaev-Zeldovich effect and suggests the existence of links between basic statistical physics and complex applications involving radiative processes. 
\end{abstract}

\section{Introduction}

One of the most relevant physical phenomena in modern cosmology is the Sunyaev-Zeldovich effect (SZE). Cosmic microwave background (CMB) photons get Comptonized when they interact with electrons in systems such as the hot gas present between clusters of galaxies. The frequency shift associated with the Compton effect cause distortions to the Planck CMB curve ($T \simeq 2.725 K$). These distortions were first quantified using the Kompaneets equation, which corresponds to a photon diffusion approach to the problem~\cite{SZ1}.

Two basic phenomena have been identified in  the SZE in terms of the motion of the electrons. The first one is related to the bulk motion of the  cluster (kinematic SZE), while the second one corresponds to the random motion of the particles present in the intracluster gas (thermal SZE). Both effects are important in order to determine cosmological parameters such as the Hubble constant, as well as primary anisotropies in the CMB spectrum \cite{ColaF1,ColaF2}. 

The photon diffusion approach was questioned by Rephaeli nearly 25 years ago, due to the low density of the gas. This author also noticed that mild-relativistic effects become relevant  to determine cluster velocities in several astrophysical scenarios \cite{SZ3}. The approach taken by Rephaeli to describe the thermal SZE was based on photon scattering techniques that involved convolution integrals that were evaluated numerically. After that work, there have been lots of contributions regarding alternative derivations of the SZE and the inclusion of other possible causes of additional CMB distortions such as magnetic fields, double Compton scattering or very large wavelength acoustic waves \cite{ASV}. More recent work suggests a relation between the SZE and the characterization of dark matter particles related to high energy reactions \cite{Lav}. 

The   present paper aims is to present a derivation of the SZE based on a simple kinetic $1D$ model of the electron gas in order to show statistical properties that link the photon scattering approach to other branches of statistical physics. The formalism reproduces both the kinematic and thermal effects in the non-relativistic regime and suggests extensions of the SZE formalism to interdisciplinary areas corresponding to low-density limits of diffusive-type processes. 

To accomplish this task, the paper has been divided as follows: In section two, the basic thermodynamic properties of black body radiation are  reviewed. Section three is dedicated to the derivation of the kinematic SZE, while section four is devoted to the analysis of the thermal SZE in which the present approach is compared with the original formalism based on the Kompaneets equation. Section five includes final remarks and a brief description of future work regarding this active area of research.  

\section{CMB Basics}

The starting point of the formalism is the occupation number for a Planck distribution:
\begin{equation}
n_{(0)}(\nu)=\frac{1} {e^{\frac{h \nu}{ k T}}-1},
\end{equation}
where $h=6.626 \times 10^{-34}$ J$ \cdot $s is the Planck constant, $k=1.38 \times 10^{-23}$ J/K is Boltzmann's constant 
and $T = 2.725$ K is the CMB temperature.
The internal energy density corresponding to the occupation number is given by
\begin{equation}
u(v)=\frac{8 \pi h \nu^3}{c^3}n_{(0)}(\nu).
\end{equation}
The intensity associated with the internal energy density reads:
\begin{equation}
I(\nu)=\frac{c}{4  \pi} u(v).
\label{eq:CMBI}
\end{equation}
In CMB physics it is customary to define de dimensionless frequency $x$ as:
\begin{equation}
x=\frac{h \nu}{ k T},
\end{equation}
Eq. (\ref{eq:CMBI}) can then be rewritten as:
\begin{equation}
I(x)=\frac{2 (k T)^3}{(h c)^2} \frac{x^3}{e^x-1}.
\label{eq:CMBx}
\end{equation}
The maximum intensity value corresponding to Eq. (5) is $x\simeq 2.821 $ which corresponds to the microwave frequency $\nu\simeq 1.601 \times 10^{11}$ Hz at the current CMB temperature.

\section{Kinematic SZE}

In the simplest model, all the electrons present in the dilute gas move with velocity $u_k$ along the $x$ direction. The dimensionless parameter $\beta_{k}$ is given by:
\begin{equation}
\beta_{k}=\frac{u_k}{c}.
\end{equation}
The frequency shift $\hat{\nu}$ for CMB photons reads:
\begin{equation}
\hat{\nu}=\nu(1  \pm \beta_{k}).
\end{equation}
Now, if $\tau$ represents the fraction of photons scattered by the electrons, the perturbed occupation number $n_{(p)}(\nu)$ is given by:
\begin{equation}
n_{(p)}(\nu)=n_{(0)}(\nu)-\tau n_{(0)}(\nu)+\tau n_{(0)}( \hat{\nu}).
\end{equation}
The last term in the right-hand side of  Eq. (8) corresponds to the number of photons with  frequency  $\hat{\nu}$ after the photon-electron interactions. This term can be approximated using a Taylor series expansion as:
\begin{equation}
n_{(0)}(\hat{\nu})\simeq n_{(0)}(\nu)\pm \beta_k \bigskip \nu \frac{\partial n_{(0)}}{\partial \nu} ,
\end{equation}
so that Eq. (8) now reads:
\begin{equation}
n_{(p)}(\nu)-n_{(0)}(\nu)=\Delta n=\pm \beta_{k} \tau \frac{\partial n_{(0)}}{\partial \nu} \nu.
\end{equation}
To compute the change in the intensity spectrum $\Delta I=I_{(p)}-I_{(0)}$ due to the kinematic SZE one can apply the relation:
\begin{equation}
\frac{\Delta I}{I_{(0)}}=\frac{\Delta n}{n_{(0)}},
\end{equation}

\vspace{10pt}

 \begin{figure}[h!]
 	\centering	
 	
 \includegraphics[width=14cm,height=7cm]{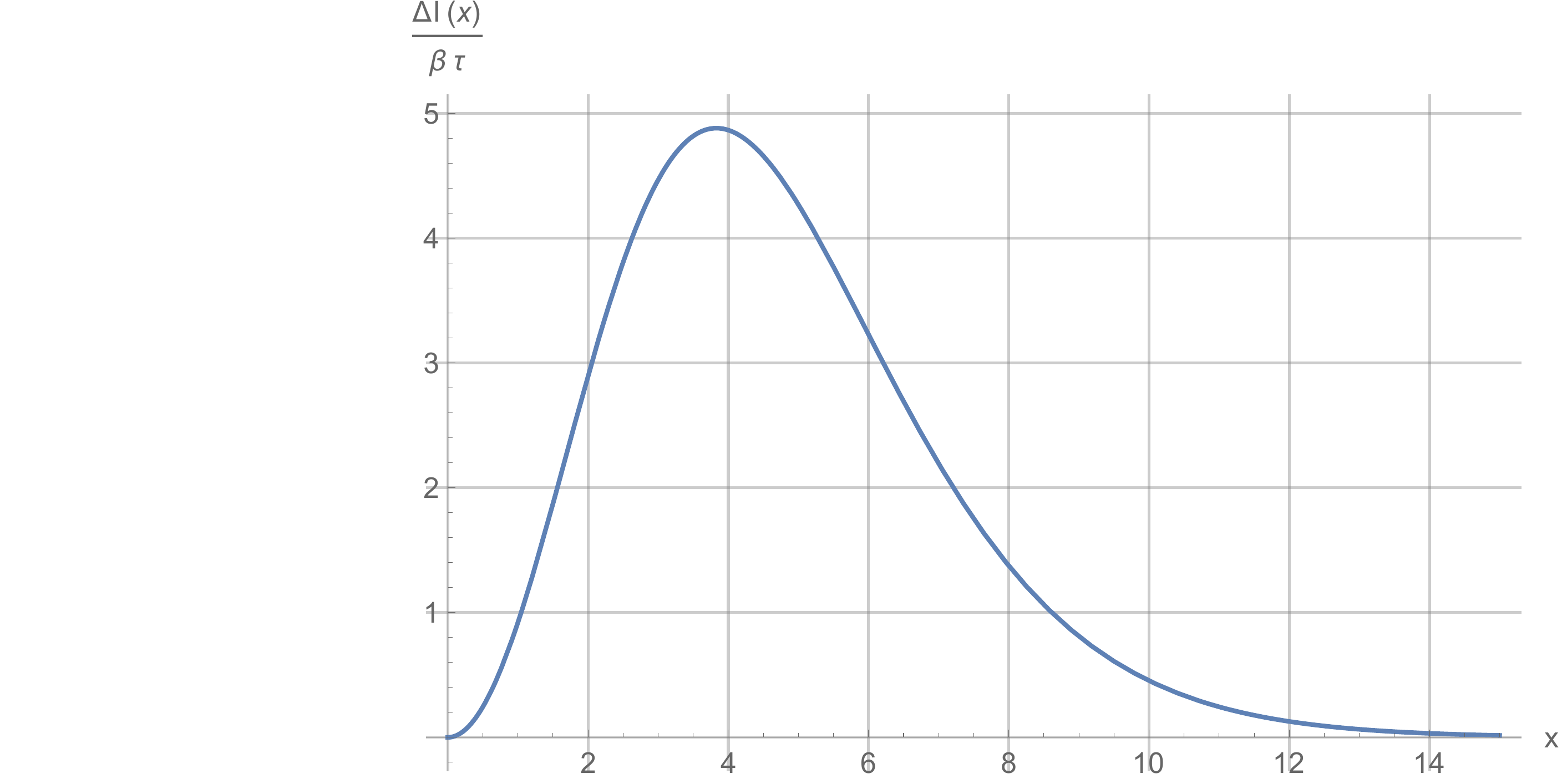}
 \caption{Kinematic SZE. The intracluster gas is assumed to move homogeneously with scaled speed $-\beta$ producing an intensity distortion. $\Delta I$ is expressed as a function of $x=\frac{h \nu}{k T}$ and in units of $\frac{2(k T)^3}{(h c)^2}$.}
 	
 \end{figure}
	
	\noindent 
	${\Delta I}/{I_{(0)}}$  can now be established using Eqs. (1), (4) and (11), thus obtaining
	\begin{equation}
	\Delta I(x)=\mp \frac{2 (k T)^3}{(h c)^2}\frac{x^4 e^x}{(e^x-1)^2} \beta_{k} \tau
	\end{equation}
	Eq. (12) is the well-known expression of the kinematic SZE \cite{SZ3}. Its shape, scaled in terms of 
	the factor $\beta \tau $ is shown  in Fig. 1.
	
	\section{Thermal SZE}
	
	We now consider the case in which the electron velocities satisfy a given $1 D$ distribution function $P=P(v)$. Photons of different frequencies contribute to the perturbed occupation number $n_{(p)}(\nu)$ in the SZE. Historically, $n_{(p)}(\nu)$ was first established using the Kompaneets equation approximation: 
	\begin{equation}
	\frac{n_p-n_{0}}{\tau z}=\frac{\Delta n(\nu)}{\tau z}= 4 x \frac{\partial n_{(0)}}{\partial x}
	+x^2  \frac{\partial^2 n_{(0)}}{\partial x^2} ,
	\end{equation}
	where $z=({k T_{el}}/{m c^2})$ is the relativistic parameter for a free electron gas at temperature $T_{el}$.
	Equation (13) can be established using a kinetic theory formalism in which the drift term, usually present 
	in the relativistic Boltzmann equation, is neglected. In that case, the photon occupation number is modified 
	due to the Compton interactions included in the corresponding collision kernel \cite{Bernstein}.
	
	  In the present formalism, the varying electron velocities are expressed as
	 \begin{equation}
	 \bar{\beta}_{Th}=\frac{v}{c} .
	 \end{equation}
	 The thermal SZE corresponds to a slight variation of Eq. (8) that reads:
	 \begin{equation}
	 n_{(p)}(\nu)=n_{(0)}(\nu)-\tau {n}_{(0)}(\nu)+\tau n_{{(d)}}(\nu),
	 \end{equation}
	 where  $n_{{(d)}}(\nu)$ is the occupation number of the photons that contribute to a fixed perturbed frequency band, which is given by:
	 \begin{equation}
	 n_{{(d)}}(\nu)=\int\limits_{-\infty}^{\infty}P(\bar{\beta}_{Th}) n(\nu+\bar{\beta}_{Th} \nu)d \bar{\beta}_{Th}.
	 \end{equation}
	 The Taylor expansion of $n(\nu+\bar{\beta}_{Th} \nu)$ up to second order in $\Delta \nu= \bar{\beta}_{Th} \nu$ leads to the expression:
	 
	 \begin{equation}
	 	n_{d}(\nu)  \simeq \int\limits_{-\infty}^{\infty}P(\bar{\beta}_{Th})( n_{(0)}+\nu \bar{\beta}_{th}\frac{\partial n_{(0)}}{\partial \nu}
	 	+\frac{\nu^2 \bar{\beta}_{th}^2}{2}\frac{\partial^2 n_{(0)}}{\partial \nu^2})d \bar{\beta}_{Th},
	 \end{equation}
	 or, using Eq. (4):
	 \begin{equation}
	 	n_{d}(x)  \simeq n_{(0)}\int\limits_{-\infty}^{\infty}P(\bar{\beta}_{Th})d \bar{\beta}_{Th}+x \frac{\partial n_{(0)}}{\partial x}
	 	\int\limits_{-\infty}^{\infty} \bar{\beta}_{Th}P(\bar{\beta}_{Th})d \bar{\beta}_{Th}
	 	+\frac{x^2}{2}\frac{\partial^2 n_{(0)}}{\partial x^2} 
	 	\int\limits_{-\infty}^{\infty} \bar{\beta}_{Th}^2 P(\bar{\beta}_{Th})d \bar{\beta}_{Th}.
	 \end{equation}
	 
	 \vspace{10pt}
	 
	 In  Eq. (18) $P(\bar{\beta}_{Th})$ is assumed to be normalized, so that 
	 \begin{equation}
	 \int\limits_{-\infty}^{\infty}P(\bar{\beta}_{Th}) d\bar{\beta}_{Th}=1.
	 \end{equation}
	 
	 The first two moments of the distribution function lead to the Kompaneets equation structure. The coefficients corresponding to the right-hand side of Eq. (13) are: 
	 \begin{equation}
	 \int\limits_{-\infty}^{\infty} \bar{\beta}_{Th} P(\bar{\beta}_{Th}) d\bar{\beta}_{Th}=4z,
	 \end{equation}
	 \begin{equation}
	 \int\limits_{-\infty}^{\infty} \bar{\beta}_{Th}^2 P(\bar{\beta}_{Th}) d\bar{\beta}_{Th}=2z.
	 \end{equation}
	 
	 The distribution function in the $1D$ model is then given by:
	 \begin{equation}
	 P(\bar{\beta}_{Th})=\frac{1}{2 \sqrt{ \pi z}} e^{-\frac{\bar{\beta}_{Th}^2}{4 z}}+\frac{\beta}{\sqrt{ \pi z}} e^{-\frac{\bar{\beta}_{Th}^2}{4 z}},
	 \end{equation}
	 Eqs. (18) and (22) lead us to the perturbed occupation number, scaled in terms of the factor $y= \tau z$:
	 \begin{equation}
	 \frac{\Delta n}{y}= \frac{x e^x(4+x+e^x(x-4))}{(e^x-1)^3}.
	 \end{equation}
	 Figure 2 shows the perturbed occupation number, Eq. (23) for several temperatures. 
	 
	  \begin{figure}[h!]
	 	\centering	
	 	
	 	\includegraphics[width=14cm,height=7cm]{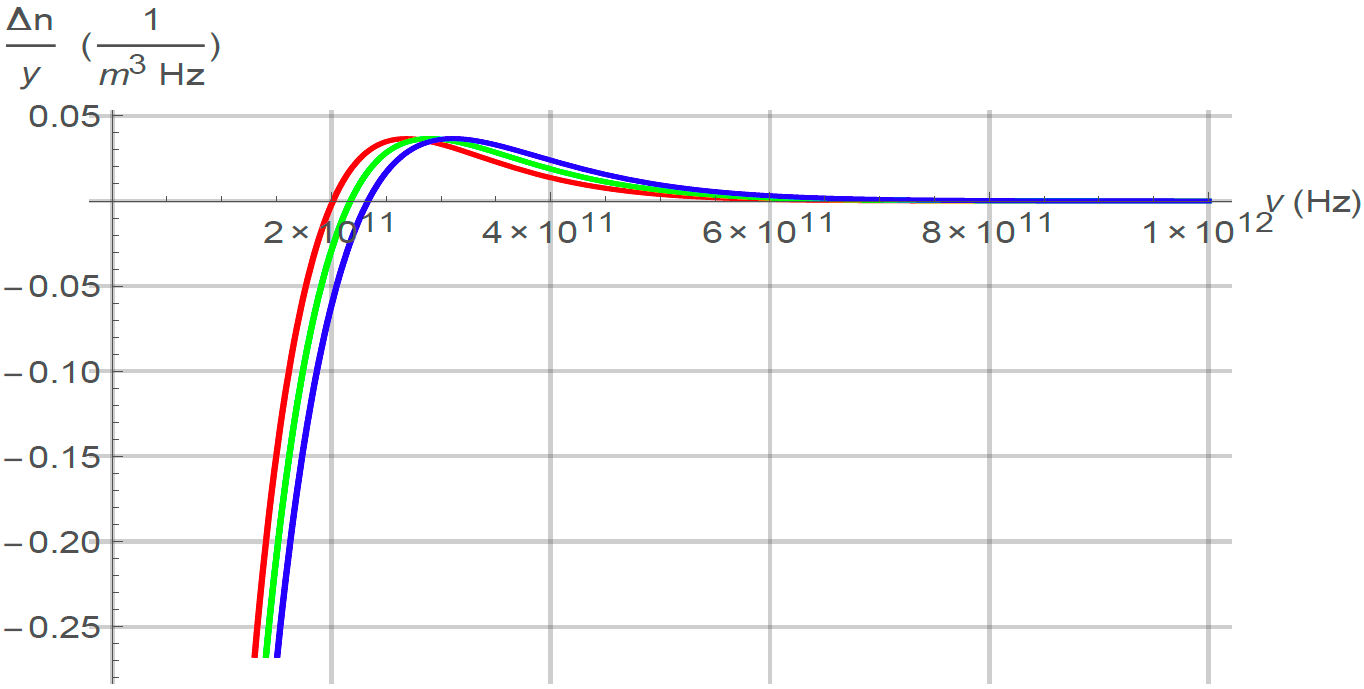}
	 	\caption{Normalized non-relativistic perturbed occupation number 
	 		$({\Delta n}/{\Delta n(\mu=0)})$ (Eq. 23). The red curve corresponds to $\mu=9 \times 10^{-5}$, 
	 		the green line to $\mu=0$, and the blue curve to $\mu=-9 \times 10^{-5}$. } 
	 	
	 \end{figure}
	 
	 At this point, Eqs. (18) and (22) can be applied in various interesting situations.   In cosmology, for example, it is known that the unperturbed occupation number (1) may be slightly modified by the presence of a non-vanishing chemical potential $\mu$, where  $  -9\times 10^{-5}\leq  \mu  \leq 9 \times 10^{-5}$ \cite{CH1}. In this case, the change in the occupation number $n_{\mu}$ reads:
	 \begin{equation}
	 \Delta n_{\mu}= y\frac{x e^{x+\mu}(4+x+e^{x+\mu}(x-4))}{(e^{x+\mu}-1)^3}.
	 \end{equation}
	 
	 Figure 3 shows the effect of this variable for $T=2.725$ K to the change of the occupation number at $\mu=0$.
	 
	 \begin{figure}[h!]
	 	\centering	
	 	
	 	\includegraphics[width=14cm,height=7cm]{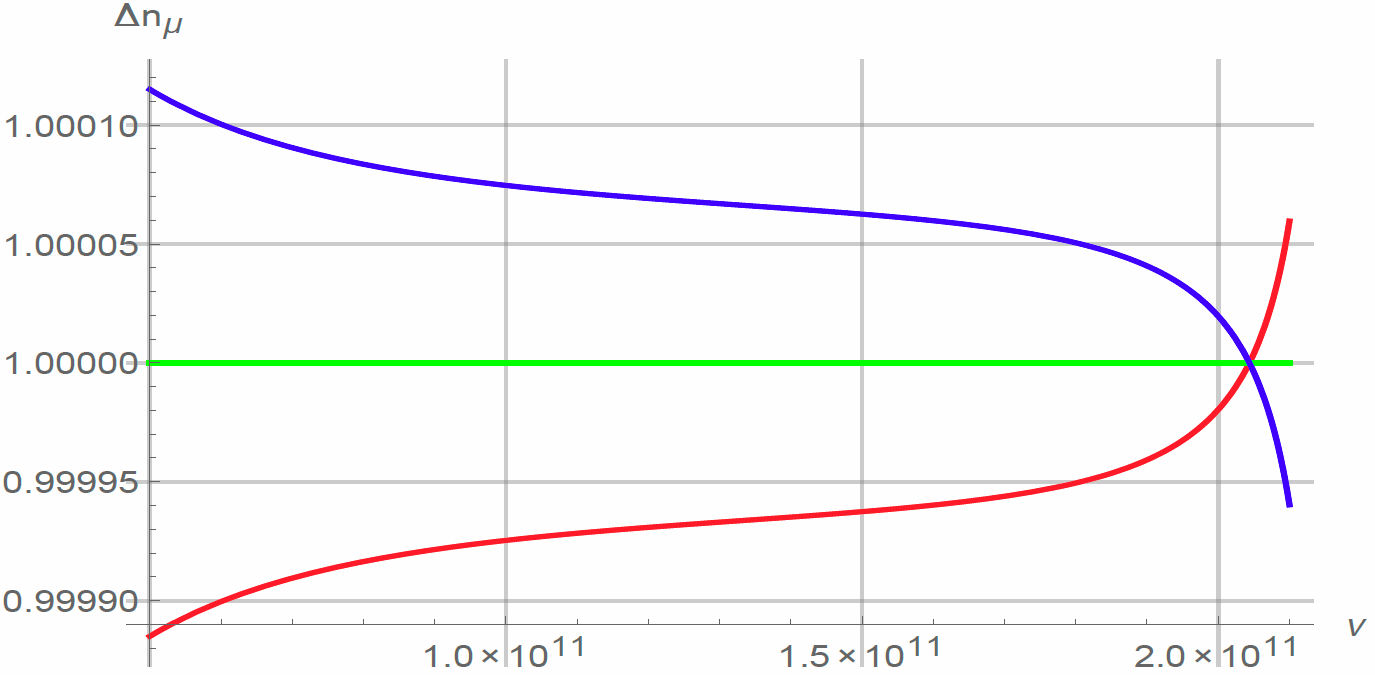}
	 	\caption{Non-relativistic change in the occupation number ${\Delta n}/{y}$ (Eq. 23). The red curve corresponds to 
	 		$T=2.525$ K, the green curve to $T=2.725$ K and the blue curve to $T=2.925$~K.} 
	 	
	 \end{figure}
	 
	 Another application of the present formalism is related to the study of other sources of distortions of the CMB curve, such as the  DC scattering \cite{Mandel}. In this case,  Eqs. (15) and (16) would require additional terms corresponding to the contributions to the band of frequency $\nu$  due to the secondary photons produced in the effect \cite{CH2}. A detailed calculation of this phenomenon is beyond the scope of the present paper, but will be the subject of future work since interdisciplinary situations in which populations are modified through "multiple births" can be described using kinetic formalisms such as the one here presented. The analogy of those kinds of situations with the one present in the DC effect seems quite promising.

	   It is also possible to apply Eq. (11)  to establish the intensity distortion curve corresponding to the thermal SZE. The  result reads:  
	  \begin{equation}
	  \frac{\Delta I(x)}{y}=\frac{2 (k T)^3}{(h c)^2}\frac{x^4 e^x}{(e^x-1)^2}(\frac{x e^x+x}{e^x-1}-4),
	  \end{equation}
	  Eq. (24) is the well-known expression for the thermal  non-relativistic SZE \cite{SZ3}. 
	  The distortion is shown in Fig. 3.

 \section{Final Remarks}
 
 This work has been devoted to the analysis of the SZE in terms of a simplified 1D kinetic model. One of the results here obtained is the establishment of a Kompaneets-type equation in which the coefficients of the derivative terms correspond to the first moments of the distribution function of the scatterers. Parity is a relevant feature to be considered in the structure of the distribution function, as was noticed in earlier work related to the SZE \cite{GC}.

  \begin{figure}[h!]
 	\centering	
 	
 	\includegraphics[width=14cm,height=7cm]{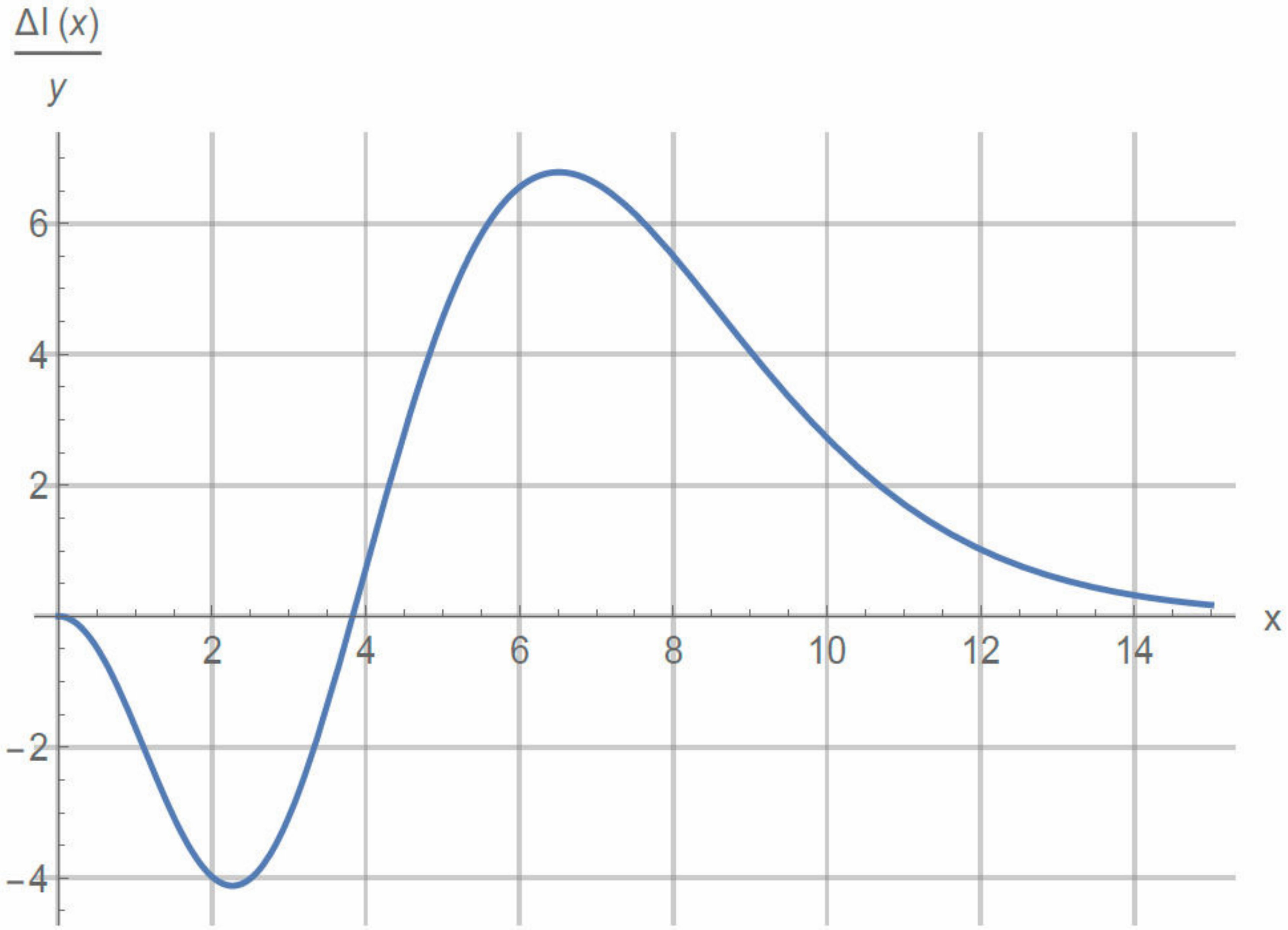}
 	\caption{ Thermal non-relativistic SZE}
 	
 \end{figure}

 \vspace{10pt}

 	It is interesting to notice that the kinematic SZE is also compatible 
 	with the simplified kinetic approach here proposed. Indeed, if 
 	$P(\bar{\beta})=\delta(\bar{\beta}-\beta)$ Eq.(16) becomes:
 	
 	\begin{equation}
 	n_{d}(x)=x \frac{\partial n_{(0)}}{\partial x}\int\limits_{-\infty}^{\infty} \bar{\beta}\delta(\bar{\beta}-\beta)d \bar{\beta},
 	\end{equation}
 	which immediately leads to Eq. (10). 
 	
 	The SZE is a 3D phenomenon, which involves several physical processes present in ionized gases. In contrast, the present approach corresponds to a simple kinetic model that suggests the existence of a direct link between the diffusive-type formalisms and the use of scattering kernels. This type of formalism allows the direct calculation of perturbed occupation numbers for other physical systems. Further applications of the type of approach here presented will include relativistic effects, as well as multiple scattering scenarios in dense systems.

	\section*{Acknowledgements}

	The author wishes to thank A. Sandoval-Rubalcava and A.R. Sagaceta-Mej\'ia for their contributions to this article. This work has been supported by the Institute of Technology and Applied Research (INIAT) of U. Iberoamericana, Mexico.

\end{document}